\documentclass[conference]{IEEEtran}
\IEEEoverridecommandlockouts
\usepackage{cite}
\usepackage{hyperref} 
\usepackage{amsmath,amssymb,amsfonts}
\usepackage{algorithmic}
\usepackage{graphicx}
\usepackage{comment}
\usepackage{textcomp}
\usepackage{xcolor}
\usepackage{subfigure}   
\def\BibTeX{{\rm B\kern-.05em{\sc i\kern-.025em b}\kern-.08em
    T\kern-.1667em\lower.7ex\hbox{E}\kern-.125emX}}
\begin{document}

\title{RespDiff: An End-to-End Multi-scale RNN Diffusion Model for Respiratory Waveform Estimation from PPG Signals\\
{}
\thanks{
\\
* Co-first Author\\
\dag Corresponding Author ym520@ic.ac.uk\\
This project is open source \url{https://github.com/MYY311/RespDiff}
}
}

\author{ \IEEEauthorblockN{Yuyang Miao$^{1\dag}$, Zehua Chen$^{1*}$, Chang Li$^{2}$, Danilo Mandic$^{3}$} \IEEEauthorblockA{$^1$ Department of Electrical and Electronic Engineering, Imperial College London, London, UK} \IEEEauthorblockA{$^1*$ Department of CST, Tsinghua University, Beijing 100084, China} \IEEEauthorblockA{$^2$ University of Science and Technology of China, Hefei, China} \IEEEauthorblockA{$^3$ Department of Electrical and Electronic Engineering, Imperial College London, London, UK}}

\maketitle

\begin{abstract}
Respiratory rate (RR) is a critical health indicator often monitored under inconvenient scenarios, limiting its practicality for continuous monitoring. Photoplethysmography (PPG) sensors, increasingly integrated into wearable devices, offer a chance to continuously estimate RR in a portable manner. In this paper, we propose RespDiff, an end-to-end multi-scale RNN diffusion model for respiratory waveform estimation from PPG signals. RespDiff does not require hand-crafted features or the exclusion of low-quality signal segments, making it suitable for real-world scenarios. The model employs multi-scale encoders, to extract features at different resolutions, and a bidirectional RNN to process PPG signals and extract respiratory waveform. Additionally, a spectral loss term is introduced to optimize the model further. Experiments conducted on the BIDMC dataset demonstrate that RespDiff outperforms notable previous works, achieving a mean absolute error (MAE) of 1.18 bpm for RR estimation while others range from 1.66 to 2.15 bpm, showing its potential for robust and accurate respiratory monitoring in real-world applications.
\end{abstract}


\section{Introduction}
Respiratory waveform monitoring is vital in clinical as it is important to indicate the health condition of patients. Respiratory rate (RR) is informative to several fatal diseases within the respiratory and cardiovascular systems \cite{gravelyn1980respiratory,strauss2014prognostic}. Hence, the methods to accurately measure RR in a convenient manner are gaining increasing attention 


Usually, respiratory signal is measured with methods such as impedance pneumography, capnography and spirometry. However, these methods require bulky machines, which are not portable and restrict the scenarios for measuring RR. Furthermore, in clinical settings, manually counting is still considered the primary technique to estimate respiratory rates. This is inconvenient for the widespread respiratory rate estimation. Thus, there is a need to extract respiratory rates or estimate respiratory waveform in a convenient and human-labour-free manner. Meanwhile, Photoplethysmography (PPG) signals have gained broad interest and have been widely integrated into wearable devices recently. PPG sensor measures the blood volume change in the microvascular bed of tissue. The PPG signals are modulated by the respiratory signals, including frequency, amplitude and baseline modulations \cite{l2019photoplethysmographic}. Then it will be beneficial to extract respiratory information from PPG signals, utilizing the portable nature of PPG sensors leading to continuous and portable RR estimation.

However, previous works show limited performance in RR estimation using deterministic mapping methods and hand-crafted features. Meanwhile diffusion model has gained tremendous interest recently due to its powerful iteratively optimizing ability \cite{ho2020denoising}. The diffusion model has shown its prominent performance in various domains, including speech generation \cite{jeong2021diff,liu2023audioldm}, time series generation \cite{rasul2021autoregressive,yuan2024diffusion,qian2024timeldm}, time series imputation \cite{tashiro2021csdi} and biomedical signal denoising \cite{liu2024sdemg}. Thus we leverage the diffusion model and introduce the diffusion model to the respiratory waveform estimation task for the first time. We also propose multi-scale encoders under the assumption that the modulation of breathing activities on PPG signals scatters over different frequency ranges. Besides, we aid the optimization of the diffusion model by adding a spectral loss term, directly measuring the sampling quality in the frequency domain. With the diffusion model's strong modelling capacity, we utilize the whole dataset without deleting signal segments due to low signal quality, which better mimics real-world scenarios.  



In summary, we propose RespDiff, an end-to-end multi-scaled RNN based diffusion model giving the key contributions

\begin{itemize}
    \item We have applied the diffusion model on the respiratory waveform extraction task for the first time, manifesting the robust performance of the diffusion model in this new field.
    \item We propose multi-scale encoders extracting spatial features at various resolutions.
    \item We integrate spectral loss into the training scheme, which significantly increases the performance of the model.
    \item We validated our method on the BIDMC dataset \cite{pimentel2016toward} and outperformed several well-established previous works, resulting in a new record in RR estimation.
\end{itemize}

\section{Related Works}
Various works have been proposed to estimate RR from PPG signals. Dong \textit{et al.} first applied classical signal processing techniques to extract multiple respiratory waveforms and remove trials with low signal quality indexes. Then they fused the extracted signals using sequence processing deep learning algorithms \cite{dong2023whale}. Iqbal \textit{et al.} removed low-quality signals and combined preprocessing, filtering and postprocessing techniques to give the final RR estimation \cite{iqbal2022photoplethysmography}. Osathitporn \textit{et al.} proposed RRWaveNet, a U-net structure Network with Convolutional Neural Networks (CNN) layers as encoders and decoders. The proposed RRWaveNet works on a processed dataset with low signal quality segments deleted to give the final RR estimation \cite{osathitporn2023rrwavenet}. Aqajari \textit{et al.} reformulated the respiratory waveform generation task as a conditional generation task and chose cycle GAN as the generation model\cite{aqajari2021end}.


    

\begin{figure*}[ht]
    \centering

    \includegraphics[width=1\textwidth]{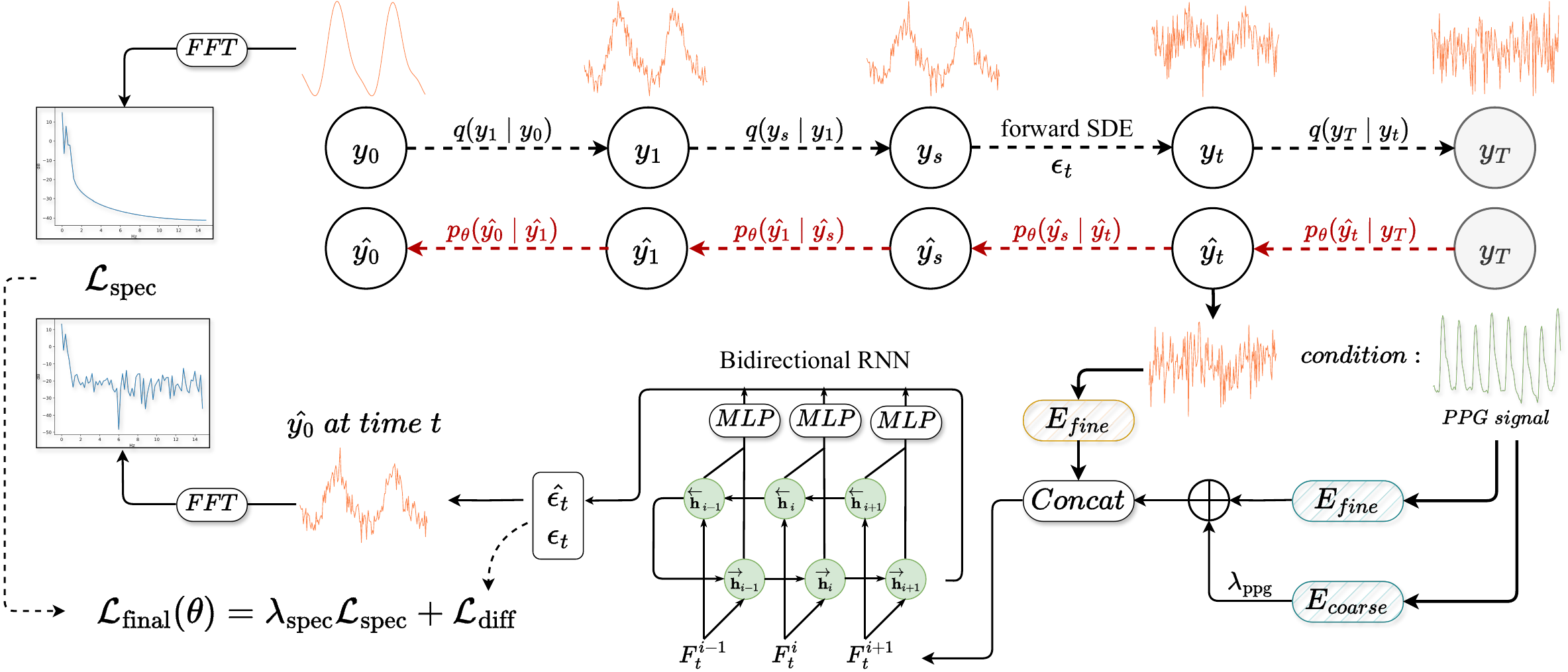}
    
    \caption{The overall architecture of the proposed RespDiff model. The respiratory waveform is gradually corrupted in the forward process and iteratively denoised in the backward diffusion. During training, features are extracted from the PPG signal and corrupted respiratory waveform to estimate the noise. Losses in the diffusion domain and frequency domain are calculated and added together.}
    \label{overall}
\end{figure*}

\section{Methods}
In this work, we have introduced a conditional diffusion model to the task of respiratory waveform estimation task. Under the backbone of the bidirectional RNN model, we propose novel multi-scale encoders which help extract features at different resolutions. Also, in addition to the diffusion loss in the distribution domain, we bring in a spectral loss term, optimizing the model regarding of signal quality in the frequency domain.

\subsection{Conditional Diffusion Models}
\label{sec:diff} 
Conditional diffusion models are composed of two processes: a forward process to corrupt the clean data to a known prior distribution, \textit{e.g.}, standard Gaussian noise, and a reverse process to iteratively recover the data distribution from the prior under the guidance of condition information.

In this work, given respiratory waveform $y \in \mathbb{R}^{L}$ and corresponding PPG signal $x_\text{PPG}\in \mathbb{R}^{L}$, we develop a waveform-domain diffusion model to generate $y$ conditioned on $x_\text{PPG}$.
In the forward process, we progressively corrupt the clean respiratory waveform $y_{0}$ into a standard Gaussian noise $y_{T} \sim \mathcal{N}(0,I)$ with a predefined noise schedule $0 < \beta_1 < \beta_2 < ... < \beta_{t-1} < \beta_T < 1$, where the transition probability can be written as:
\begin{equation}
    q(y_t \mid y_{t-1}) = \mathcal{N}(y_t; \sqrt{1 - \beta_t}y_{t-1}, \beta_tI).
    \label{forward_1}
\end{equation}
With the property of isotropic Gaussian noise, we can efficiently calculate the noisy representation at time step $t$ with:
\begin{equation}
    q(y_t \mid y_0) = \mathcal{N}(y_t; \sqrt{\Bar{\alpha_t}}y_0, (1-\Bar{\alpha_t})\epsilon),
    \label{forward_2}
\end{equation}
where $\alpha_t = 1 - \beta_t$ and $\Bar{\alpha_t} = \prod_{s=1}^{t}\alpha_s$ indicate noise level and $\epsilon \sim \mathcal{N}(0,I)$ denotes the Gaussian noise injected to corrupt the clean signal $y_{0}$.

In reverse process, we start the denoising process from $y_{T} \sim \mathcal{N}(0,I)$, and gradually removes the noise added to the clean signal $y_0$ at each time step:
\begin{equation}
    p_{\theta}(y_{0:T}) = p(y_T) \prod_{t=1}^{T}p_\theta(y_{t-1} \mid y_t, x_\text{ppg}).
\end{equation}

The training objective of diffusion models is to maximize the variation lower bound (ELBO) of the likelihood of $p_{\theta}(y_0)$ \cite{ho2020denoising}. In practice, we adopt a re-weighted loss function from previous works \cite{ho2020denoising,kong2020diffwave,chen2020wavegrad} as follows:
\begin{equation}
\label{loss}
    L_\text{diff}(\theta) = \mathbb{E}_{y_0, \epsilon, t} \left\| \epsilon - \epsilon_{\theta} \left( \sqrt{\bar{\alpha}_t} y_0 + \sqrt{1 - \bar{\alpha}_t} \epsilon, t, x_\text{ppg}\right) \right\|_2^2.
 \end{equation}

In bio-electrical signal processing, several previous works have established strong diffusion baselines for EMG synthesis \cite{xiong2024patchemg} and denoising \cite{liu2024sdemg}, EEG imagine speech decoding \cite{kim2023diff}, and ECG imputation and forecasting \cite{jenkins2023improving,alcaraz2022diffusion}. However, none of the previous studies explored the performance of diffusion models for respiratory waveform, which is a vital sign of health conditions in clinical.

\subsection{Network Architecture}

\subsubsection{Multi-scale fine-grained feature encoder}
CNN is known for its powerful potential in feature extraction. The size of the kernel of CNN decides the spatial resolution of the extracted features. It is considered to be difficult to carefully design kernel sizes without any prior knowledge. To cope with this challenge, we propose the encoder $E_\text{fine}$ which aims to extract characteristics from PPG at multiple scales. The  $E_\text{fine}$ is composed of a stack of $K$ convolutional layers with varying resolution sizes. Multi-scale fine-grained feature extractor can be formulated as:
\begin{equation}
E_{\text{fine}}(x) = \text{Concat}_{i=1}^{K} \left( \text{Conv1D}(x, \text{kernel size} = k_i) \right).
\label{diffusion_equtation}
\end{equation}

\subsubsection{Multi-scale coarse-grained feature encoder} Breathing activities can generate slow-varying components in PPG signals, such as baseline modulation. Even a large kernel size is hard to capture such information. Thus we need feature extraction methods with large receptive fields. We propose the dilated multi-scale encoder $E_{coarse}$, which aims to capture coarse-grained spatial features from the PPG signal and uses a similar multi-resolution design with $E_\text{fine}$ with $N$ layers. However, the convolutional layer is replaced by dilated convolutions $\text{DilConv1D}(:, k_i)$, which significantly helps increasing the receptive field:
\begin{equation}
E_{\text{coarse}}(x) = \text{Concat}_{i=1}^{K} \left( \text{DilConv1D}(x, \text{kernel size} = k_i) \right).
\end{equation}

\subsubsection{Bidirectional RNN}
During the respiratory estimation process, the beginning of the sequence usually suffers from bad quality due to the lack of context before it. Thus, we utilize bidirectional RNN, which leverages both past and future context to improve the quality of predictions even at the beginning of the sequence. 
To make better use of the information in the input signal, firstly, the multi-scale PPG feature $f_\text{ppg}$ with feature fusion ratio $\lambda_\text{ppg} $ can be formulated as:
\begin{equation}
    f_\text{ppg} =  E_{\text{fine}}(x_\text{ppg}) +  \lambda_\text{ppg} E_{\text{corase}}(x_\text{ppg}).
\end{equation}
The bidirectional RNN processes the input feature $f_\text{ppg}$ and aims to estimate the noise illustrated in \ref{loss}. Through training, the timestep for the diffusion process is uniformly sampled with $t \sim U(0, 1)$. Instead of directly sending the noised input $y_t$ into the noise predictor, we extract its fine-grained feature $f_{y_t} = E_\text{fine}(y_t)$ and concat it with $f_\text{ppg}$ along the channel.
The fused input, denoted as $f_\text{input} = F_{t}= \text{Concat}(f_\text{ppg}, f_{y_t}) \in \mathbb{R}^{L \times d}$, is sent to a bidirectional RNN to predict the noise at state $t$. The hidden state of the forward (fd) and backward (bd) RNN process is created as:
\begin{align}
    \overrightarrow{h}^{i}_{t}  = \phi \left(F^i_{t} W_{dh}^{\text{fd}} + \overrightarrow{h}^{i-1}_{t} W_{hh}^{\text{fd}} + b_h^{\text{fd}} \right),\\
    \overleftarrow{h}^{i}_{t} = \phi \left( F^{i}_{t} W_{dh}^{\text{bd}} + \overleftarrow{h}^{i+1}_{t} W_{hh}^{\text{bd}} + b_h^{\text{bd}} \right),
\end{align}
where the weights in RNN are $W_{dh}^{\text{fd}} \in \mathbb{R}^{d \times h}$ , $ W_{hh}^{\text{fd}} \in \mathbb{R}^{h \times h}$ , $ W_{dh}^{\text{bd}} \in \mathbb{R}^{d \times h}$ , $ W_{hh}^{\text{bd}} \in \mathbb{R}^{h \times h}$, $b_h^{\text{fd}} \in \mathbb{R}^{1 \times h}$ and $b_h^{\text{bd}} \in \mathbb{R}^{1 \times h}$, and $\phi$ is the activation function.
$\overrightarrow{h}^i_{t}$ and $\overleftarrow{h}^i_{t}$, are the forward and backward hidden features at signal position $i$ and diffusion timestep $t$. The concatenation of them  $h^i_{t} = [\overrightarrow{h}^i_{t}; \overleftarrow{h}^i_{t}] \in \mathbb{R}^{1 \times 2h}$ will go through output head to give the estimated noise.




\subsection{Spectral Loss}

The training objective of diffusion models shown in~\eqref{loss} is to faithfully recover the data distribution from the prior distribution, while it may not guarantee the optimal sample quality in RR estimation.
Hence, to strengthen the sample quality of our proposed RespDiff, we further investigate the function of auxiliary loss functions which highlights the RR information in synthesized respiratory signals.
Specifically, we introduce a spectral loss into the original training objective of diffusion models.
At each training iteration, given the noisy representation $y_t$, the network predicts the added noise $\epsilon_t$ and then we can estimate a coarse respiratory waveform $\hat{y_0}$ with a single step:

\begin{equation}
    \hat{y}_0 = \frac{1}{\sqrt{\bar{\alpha}_t}}(y_t - \sqrt{1 - \bar{\alpha}_t}\epsilon_t).
\end{equation}
Then, we apply the Fourier transform magnitude extractor $\mathcal{F}_\text{fft}$ to both estimated $\hat{y}_0$ and the ground-truth signal $y_0$, and calculate a distance between them with:
\begin{equation}
    \mathcal{L}_\text{spec} = \frac{1}{N} \sum_{i=1}^{N} || \mathcal{F}_{\text{fft}}^i(\hat{y_0}) - \mathcal{F}_{\text{fft}}^i(y_0) ||^2,
\end{equation}

where $N$ is the number of frequency bins. Hence, during the training process, we employ both the diffusion loss shown in~\eqref{loss} and a weighted $\mathcal{L}_\text{spec}$, where the weight $\lambda_{spec}$ is set as 0.01 in our experiments.
In inference, at each sampling step, the generation of diffusion models has been strengthened by our proposed auxiliary loss, leading to a final improvement in RR estimation after iterative sampling steps.

\section{Experiments}

\subsection{Experimental Setup}

Our conduct experiments on the BIDMC dataset \cite{pimentel2016toward}, a widely-used benchmark dataset containing 53 recordings of ECG, PPG, and impedance pneumography signals, each of which has a length of 8 minutes.  
The PPG signal and respiratory waveform are first downsampled to 30Hz since the breathing activity predominantly happens in the low frequency domain. Then, the breathing and PPG signals are further low-pass filtered with a cutoff frequency of 1Hz, segmented into lengths of 5 seconds, and normalized to a range of [-1, 1] and [0, 1] respectively. To mimic the real-world scenarios, we retain each processed sample, rather than deleting low-quality ones~\cite{osathitporn2023rrwavenet,iqbal2022photoplethysmography}. Leave-one-subject-out method is chosen as the training scheme for a comprehensive evaluation.

At the inference stage, we employ both DDPM \cite{ho2020denoising} and DDIM~\cite{song2020denoising} sampler to generate 5-second samples and concatenate the samples to obtain 8-minute results. RR estimation error and waveform estimation error are calculated on a window size of 60 seconds, where RR is calculated by applying a Fourier transform on the waveform and then finding the maximum non-zero frequency component. Following previous works~\cite{aqajari2021end,iqbal2022photoplethysmography,osathitporn2023rrwavenet}, Mean absolute error (MAE) is used as the evaluation metric for both tasks.

\subsection{Results Analysis}
As shown in table \ref{bench}, we compare our works with three previous works, where our RespDiff outperforms other methods in RR estimation by a large margin. 
Especially, compared with RRWaveNet \cite{osathitporn2023rrwavenet} and the work from Iqbal \textit{et al.} \cite{iqbal2022photoplethysmography} which only estimate RR, our work generates the whole respiratory waveform. We perform a more complex task but achieve stronger performance. 
Moreover, our end-to-end network does not require laborious task-specific tuning, \textit{e.g.}, the threshold selection process required by the baseline Iqbal \textit{et al.} \cite{iqbal2022photoplethysmography}.

Also, our work utilizes the whole dataset without removing any segments. This approach better mimics the real-world scenario, where a clean recording environment is not a guarantee. The waveform MAE estimation results for RespDiff with and without spectral loss are 0.307 and 0.316. All mentioned works did not mention waveform estimation loss.

\begin{table}[htbp]
\caption{Benchmarking table containing our work and other recent work. * denotes that part of the data has been removed according to signal quality.}
\begin{center}
\begin{tabular}{l|c c}
\hline
\textbf{Model}&\textbf{Window Size}&\textbf{RR Error (bpm)} \\
\hline
\textbf{RRWaveNet\textsuperscript{*} \cite{osathitporn2023rrwavenet}} & 64s & 1.66 \\
\textbf{CycleGAN \cite{aqajari2021end}} & 60s & 1.90 \\
\textbf{Iqbal \textit{et al.} \cite{iqbal2022photoplethysmography}\textsuperscript{*}}& 90s & 2.05 \\
\textbf{Iqbal \textit{et al.} \cite{iqbal2022photoplethysmography}} & 90s & 2.15 \\
\textbf{RespDiff}& 60s & \textbf{1.18} \\
\hline
\end{tabular}
\end{center}
\label{bench}
\end{table}
\subsection{Ablation Study on Spectral Loss}
\begin{figure}[ht]
    \centering
    \subfigure[]{%
        \includegraphics[width=0.23\textwidth]{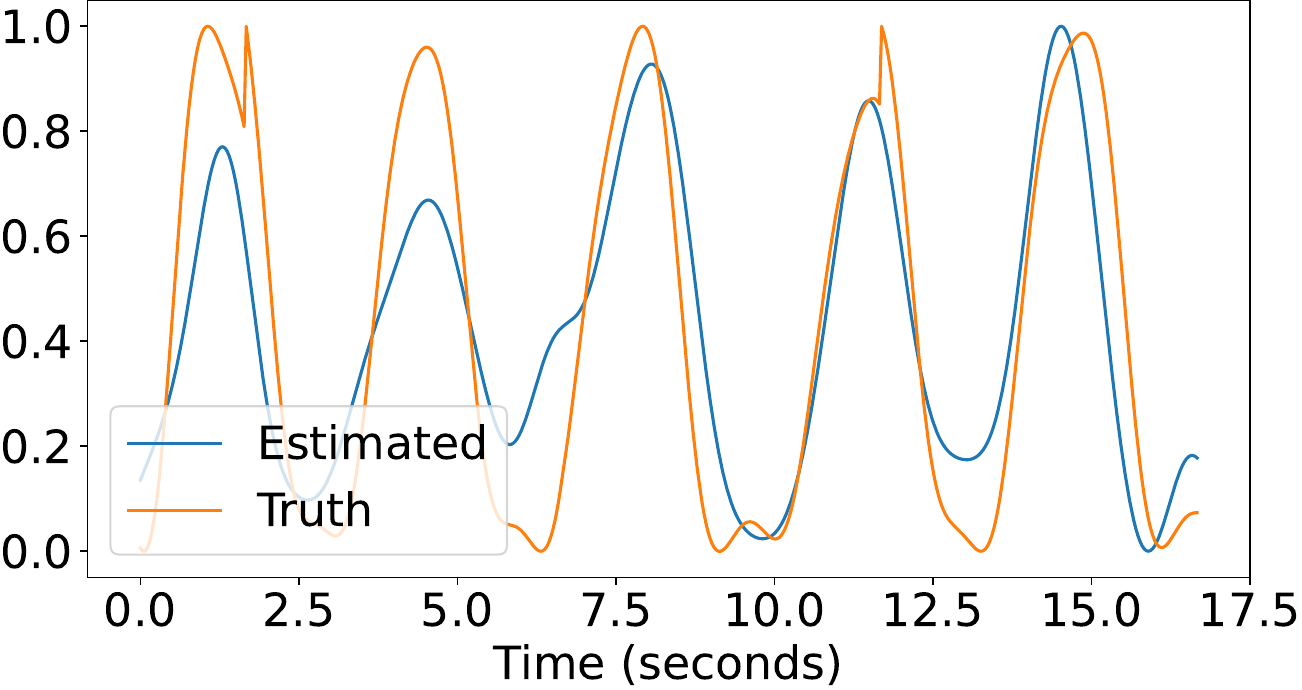}
        \label{fig:subfig_a}
    }
    \hfill
    \subfigure[]{%
        \includegraphics[width=0.23\textwidth]{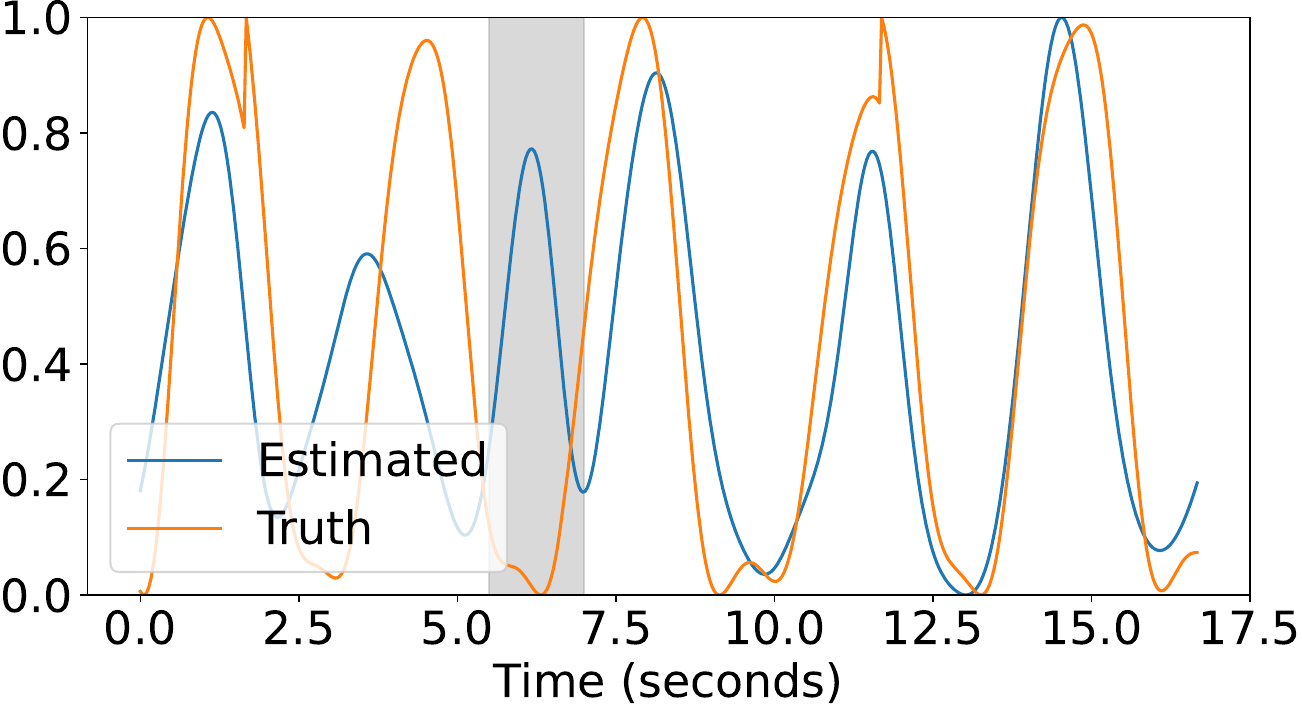}
        \label{fig:subfig_b}
    }
    \vfill
    \subfigure[]{%
        \includegraphics[width=0.23\textwidth]{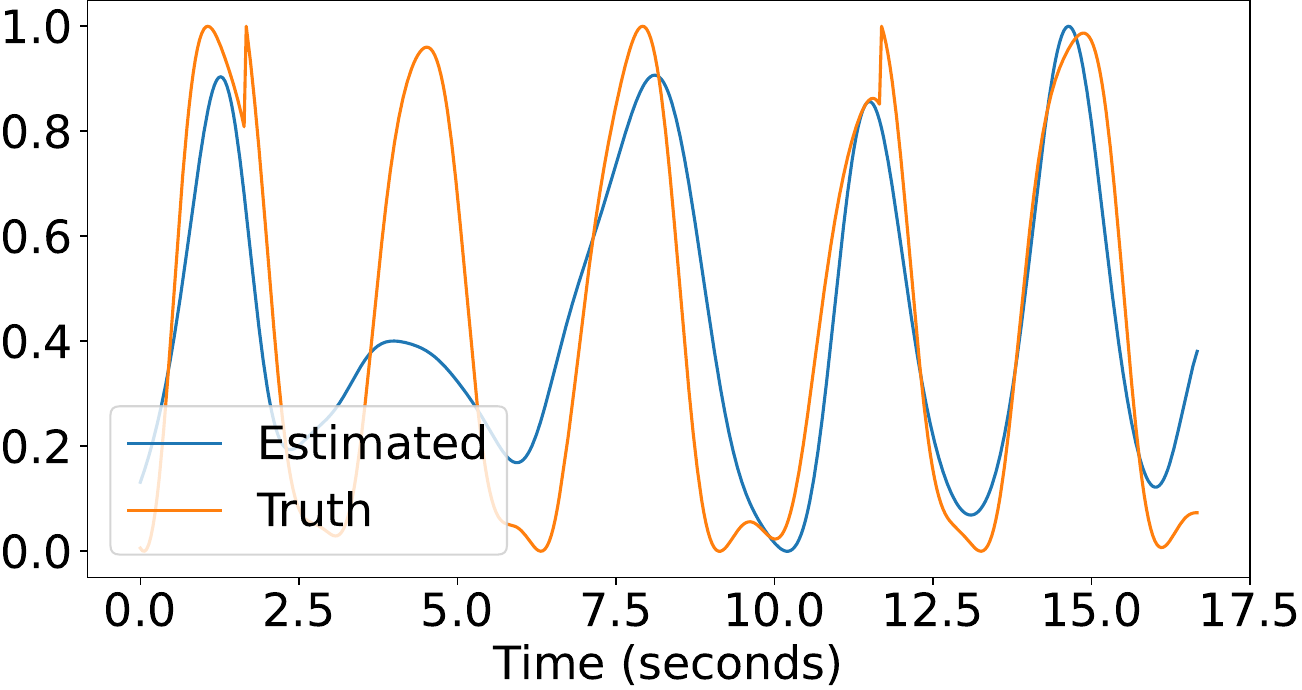}
        \label{fig:subfig_b}
    }
    \hfill
    \subfigure[]{%
        \includegraphics[width=0.23\textwidth]{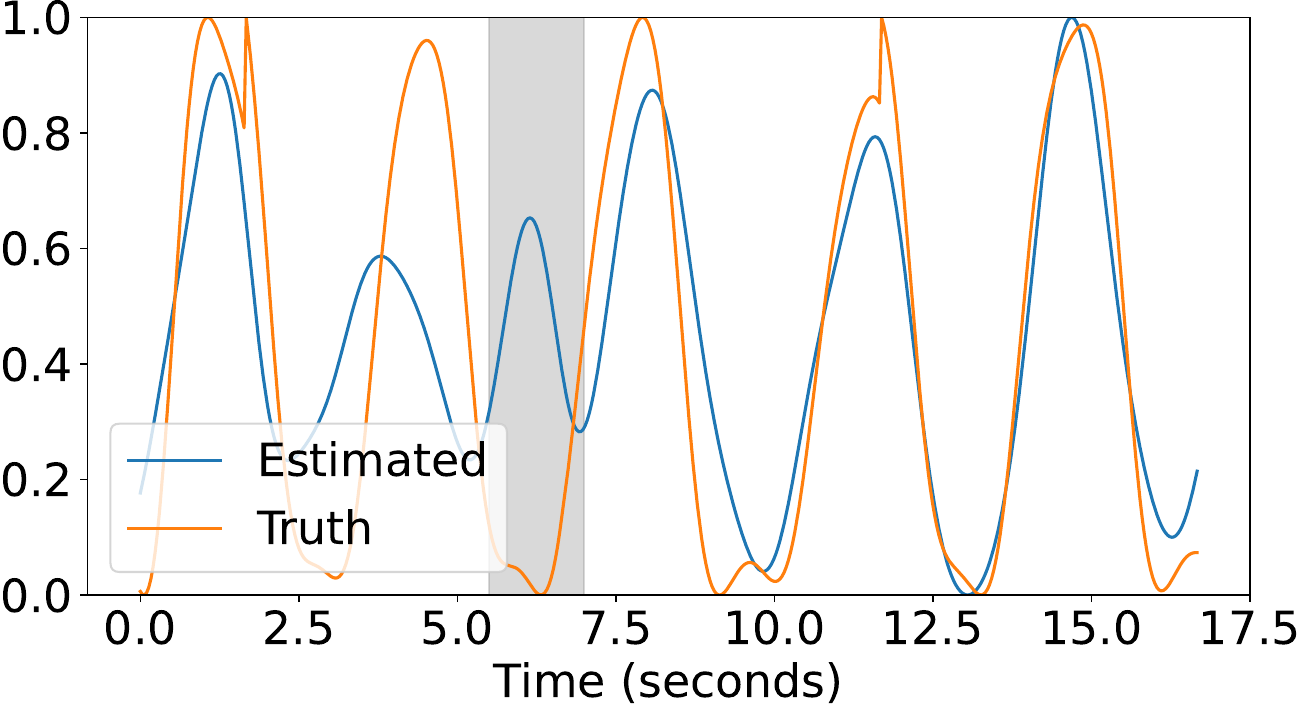}
        \label{fig:subfig_b}
    }
    \caption{Examples of respiratory waveform estimation. (a): RespDiff with 50 diffusion steps and spectral loss. (b): RespDiff with 50 diffusion steps and without spectral loss. (c): RespDiff with 6 diffusion steps and spectral loss. (d): RespDiff with 6 diffusion steps and without spectral loss.}
    \label{fig:case}
\end{figure}

We test the function of our proposed spectral loss under diverse inference processes of diffusion models, \textit{i.e.}, numerous-step sampling and few-step sampling.
As shown in Table~\ref{ddim}, in 50-step synthesis, the RR estimation error has distinctively decreased because of spectral loss.
When reducing the number of sampling steps to $6$, our proposed spectral loss still plays an important role in improving RR estimation accuracy. Notably, the spectral loss makes 6-step sampling outperform the 50-step RespDiff without this auxiliary loss, considerably improving the inference speed of RespDiff.
In our observation, with 6 inference steps, RespDiff could generate an 8-minute respiratory waveform in 7 seconds. Figure \ref{fig:case} gives an example of respiratory waveform estimation under different settings. It is manifest that when the spectral loss is not employed, the estimated respiratory waveform gives the wrong RR by generating one more peak indicated by grey shades.


\begin{table}[htbp]
\caption{Ablation on spectral loss under the different number of function evaluation (NFE).}
\begin{center}
\begin{tabular}{l c c c}
\hline
\textbf{Model}&\textbf{NFE}&\textbf{Window Size}&\textbf{RR Error (bpm)}\\
\hline
\textbf{RespDiff} &  50 & 60s & 1.18\\
\textbf{w/o spectral loss} & 50 & 60s & 1.44\\
\textbf{RespDiff} & 6 & 60s & 1.30\\
\textbf{w/o spectral loss} & 6 & 60s & 1.46\\
\hline
\end{tabular}
\label{tab1}
\end{center}
\label{ddim}
\end{table}

\subsection{Ablation Study on Multi-Scale Encoder}
To further justify the idea of applying a multi-scale encoder to extract features with different spatial resolutions. We set the kernel sizes to be all 3 instead of 1, 3, 5, 7, 9 and 11 in the multi-scale setting. Under the same training setup, the RR error increases from 1.44 bpm to 1.53 bpm, manifesting the effects of the multi-scale encoder. 

\section{Conclusion}
In this work, we have introduced RespDiff, a novel multi-scale diffusion model designed for the challenging task of respiratory waveform estimation from PPG signals. Our model leverages a bidirectional RNN as its backbone, coupled with multi-scale encoders that effectively capture features across various temporal resolutions. RespDiff has demonstrated promising results, outperforming several recent approaches in terms of both respiratory rate (RR) estimation and waveform reconstruction accuracy.

Furthermore, we have empirically shown that incorporating a spectral loss term significantly enhances the model's RR estimation capabilities, regardless of whether DDPM or DDIM sampling is employed. This highlights the importance of considering both time-domain and frequency-domain information for accurate RR estimation.

To the best of our knowledge, this is the first application of diffusion models to the problem of respiratory waveform estimation from PPG signals. Our results underscore the powerful modelling capacity of diffusion models in this new domain, opening up exciting possibilities for future research.

\bibliographystyle{IEEEtran} 
\bibliography{ref}

\end{document}